\documentclass[prl,superscriptaddress,twocolumn,showpacs,amsmath,amssymb]{revtex4-1}
\usepackage{graphicx,bm,dcolumn}
\usepackage{multirow}

\begin{document}
\title{Buried double CuO chains in YBa$_2$Cu$_4$O$_8$ uncovered by nano-ARPES}

\author{Hideaki Iwasawa}
\thanks{Correspondence and requests for materials should be addressed to H.I. (h-iwasawa@hiroshima-u.ac.jp)}

\affiliation{Diamond Light Source, Harwell Science and Innovation Campus, Didcot OX11 0DE, UK}
\affiliation{Graduate School of Science, Hiroshima University, Hiroshima 739-8526, Japan}

\author{Pavel Dudin}
\affiliation{Diamond Light Source, Harwell Science and Innovation Campus, Didcot OX11 0DE, UK}

\author{Kyosuke Inui}
\author{Takahiko Masui}
\affiliation{Department of Physics, Kindai University, Higashi-Osaka 577-8502, Japan}

\author{Timur K. Kim}
\author{Cephise Cacho}
\affiliation{Diamond Light Source, Harwell Science and Innovation Campus, Didcot OX11 0DE, UK}

\author{Moritz Hoesch}
\affiliation{Diamond Light Source, Harwell Science and Innovation Campus, Didcot OX11 0DE, UK}
\affiliation{Deutsches Elektronen-Synchrotron (DESY), Photon Science, Hamburg 22607, Germany}

\begin{abstract}
The electron dynamics in the CuO chains has been elusive in Y-Ba-Cu-O cuprate systems by means of standard angle-resolved photoemission spectroscopy (ARPES); 
cleaved sample exhibits areas terminated by both CuO-chain or BaO layers, and the size of a typical beam results in ARPES signals that are superposed from both terminations. 
Here, we employ spatially-resolved ARPES with submicrometric beam (nano-ARPES) to reveal the surface-termination-dependent electronic structures of the double CuO chains in YBa$_2$Cu$_4$O$_8$.
We present the first observation of sharp metallic dispersions and Fermi surfaces of the double CuO chains buried underneath the CuO$_2$-plane block on the BaO terminated surface. 
While the observed Fermi surfaces of the CuO chains are highly one-dimensional, 
the electrons in the CuO-chains do not undergo significant electron correlations and no signature of a Tomonaga-Luttinger liquid nor a marginal Fermi liquid is found. 
Our works represent an important experimental step toward understanding of the charge dynamics and provides a starting basis for modelling the high-$T_{\rm c}$ superconductivity in YBCO cuprate systems.
\end{abstract}

\maketitle

Identifying the role of the CuO chains is fundamentally important to understand superconductivity in high-$T_{\rm c}$ yttrium-barium-copper-oxide (YBCO) superconductors, 
in which quasi-one-dimensional (1D) CuO chains and two-dimensional (2D) CuO$_2$ planes coexist. 
Due to the low dimensional electronic nature of the CuO chains, the electrons are generally expected to be strongly correlated because their motion is confined along the atomic chains \cite{Voit95}. 
Such electron dynamics in the coupled CuO chains was calculated using a variation of the dynamical mean field theory (DMFT) where all the 1D chains except for one-site are treated as a self-consistent bath (ch-DMFT), 
demonstrating a possible splitting of the 1D Fermi surface into Fermi pockets due to diverging self-energy in the chains \cite{Berthod95}. 

Many experimental efforts using angle-resolved photoemission spectroscopy (ARPES) have been devoted on the single CuO-chain system, YBa$_2$Cu$_3$O$_{7-\delta}$ (Y123), 
while the results and interpretations of the electronic structure has been not straightforward \cite{Schabel98,Lu01,Nakayama07,Zabolotnyy07,Hossain08,Okawa09,Fournier10}. 
This is mainly due to the coexistence of multiple CuO- and BaO-terminated surfaces on the cleaved surface \cite{Edwards92,Iwasawa18}. 
On the other hand, Kondo {\it et al.} succeeded in disentangling complex electronic structures originated in the CuO-chains and CuO$_2$ planes in the double CuO-chains system, YBa$_2$Cu$_4$O$_8$ (Y124), 
by utilizing spatially-resolved ARPES with sub-hundred micrometer beam (micro-ARPES) \cite{Kondo07,Kondo09,Kondo10}. 
It is noteworthy that they observed a spectral weight distribution at the Fermi surface that resembles those predicted by ch-DMFT calculations \cite{Kondo10}, suggesting strong electron correlations in the CuO chains. 

These micro-ARPES results demonstrated that one of the CuO chains exposed on the CuO-terminated surface is insulating, while the other (subsurface) CuO chain is conducting. 
Although the mechanism of the electronic reconstruction resulting in the insulating chains is unknown, 
the reported CuO-chain states are representative of surface properties and distinct from those in the bulk, 
in which they are conductive and contribute to superconductivity \cite{Basov95,Hussey97}. 
Consequently, the electronic states of the 1D CuO chains related to the bulk superconducting properties has been not clarified yet.

In this Letter, we present the state-of-the-art ARPES study utilizing sub-micrometer probe spot (nano-ARPES) to clarify the termination-dependent electronic structure of the CuO chains in Y124. 
We observe the metallic dispersions of double CuO chains, buried underneath the CuO$_2$ plane on the BaO-terminated surface, for the first time. 
The metallic behaviour of these double CuO chains is consistent with the bulk transport properties \cite{Basov95,Hussey97}, 
and quite different from the surface reconstructed structures observed in the CuO-terminated surface as in the previous micro-ARPES reports \cite{Kondo07,Kondo10}. 
In addition, we observe a highly 1D Fermi surfaces with a tiny warping due to the interchain coupling, 
while seeing no evidence of strong electron correlations nor characteristic 1D electronic behaviour. 
Our experimental data thus show the intrinsic electronic properties of the native CuO chains of Y124 in full.

\begin{figure*}[t]
\includegraphics[width=170mm,keepaspectratio]{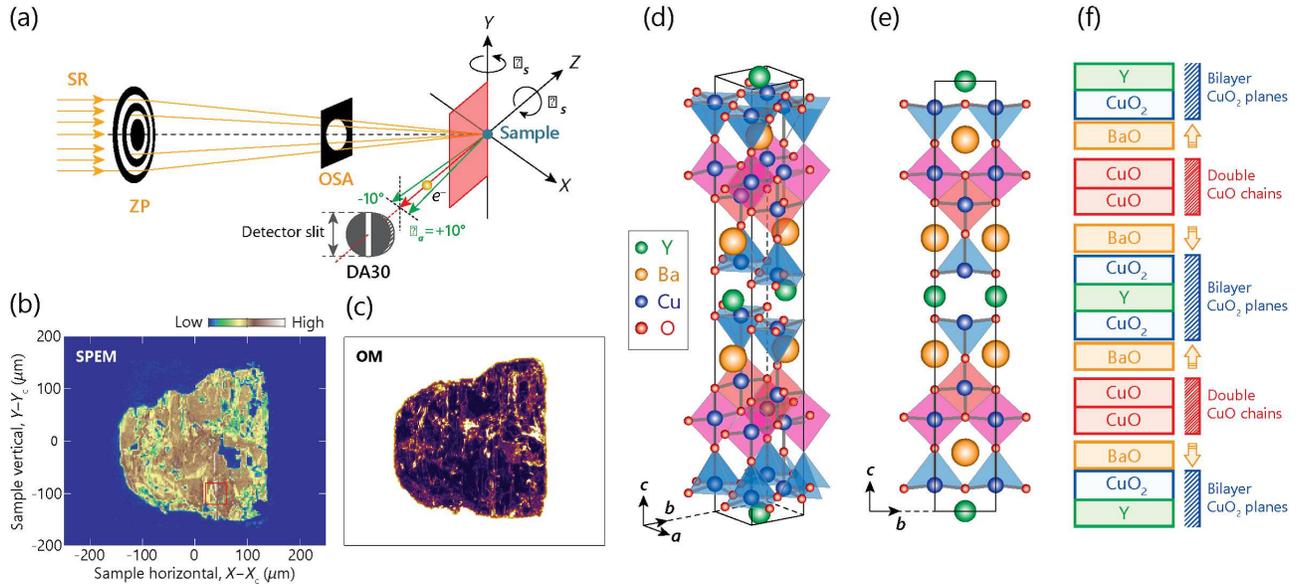}
\caption{
Overview of the nano-ARPES experiments on YBa$_2$Cu$_4$O$_8$ (Y124). 
(a) Schematic drawing of the experimental layout, where a submicrometric beam can be provided via focusing optics composed of a Fresnel zone plate (ZP) and an order sorting aperture (OSA) in close vicinity of the sample.
A deflector scan ($\theta_a$) is used for polar angular scans against the vertical detector slit, besides the manipulation of the polar ($\theta_s$) and azimuthal angles ($\varphi_s$) of the sample using goniometer.
(b) Scanning photoemission microscopy (SPEM) image taken with a one-micron step both for horizontal and vertical sample axes ($X$ and $Y$), rescaled by the sample centres ($X_c$ and $Y_c$). 
The red rectangle indicates a region of interest zoomed in Fig.~\ref{Fig2}. 
(c) Optical microscope (OM) image taken after measurements {\it ex-situ}.
(d)-(f) Orthorhombic crystal structure of Y124 with space group {\it Ammm}; 
(d) an aerial view, (e) a cross-sectional view, and (f) schematic illustration for the $c$-axis stacking of double CuO chains and bilayer CuO$_2$ planes. 
Either CuO or BaO terminated surfaces are generally expected to be exposed as a cleavage plane of the Y124 crystal.}
\label{Fig1}
\end{figure*}

\begin{figure}[t]
\includegraphics[width=85mm,keepaspectratio]{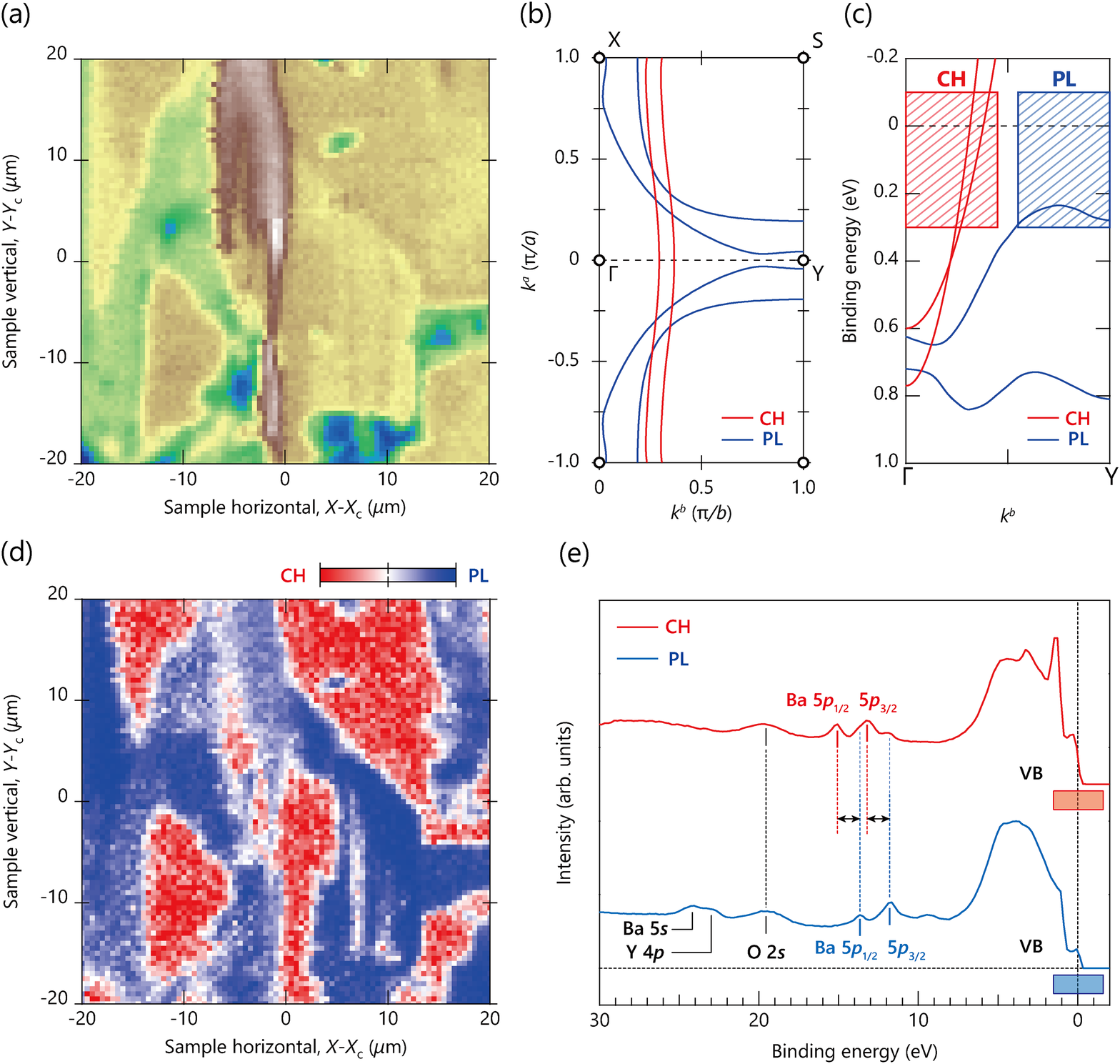}
\caption{
Characterization of surface terminations on Y124.
(a) Zoomed in spatial mapping images constructed by full integration windows in energy and momentum, taken with 500~nm step both for horizontal and vertical axes ($X$ and $Y$), rescaled by the centres of the image ($X_c$ and $Y_c$).
(b) Schematic Fermi surfaces derived from CuO$_2$ planes (blue) and CuO chain (red).
(c) Band dispersions along the high symmetry $\Gamma$Y line based on band-structure calculations \cite{Yu91}.
(d) Spatial distribution of CuO-chains and CuO-planes dominant regions (CH and PL) as indicated by red and blue regions, respectively. 
The map was obtained by $(I_{\rm CH}-\overline{I_{\rm CH}})/(I_{\rm PL}-\overline{I_{\rm PL}})$ where $I_{\rm CH}$ and $I_{\rm PL}$ represents the integrated intensity sensitive to the CuO chains and CuO$_2$ planes, respectively, 
and they were obtained by applying limited integration windows as indicated by the red and blue shaded regions in the panel~(c),
and $\overline{I_{\rm CH}}$ and $\overline{I_{\rm PL}}$ are the mean value of $I_{\rm CH}$ and $I_{\rm PL}$, respectively. 
(e) Representative termination-dependent energy distribution curves (EDCs) measured on the CH (red) and PL (blue) terminated surfaces.}
\label{Fig2}
\end{figure}

High-quality single crystals of Y124 ($T_c$ = 82~K) were grown by the flux method at ambient pressure \cite{Song07}. 
Experiments were performed using the newly-developed nano-ARPES instrument at beamline I05 of the Diamond Light Source with a spatial resolution better than 500~nm (see the Supplementary Note~1 and Note~2 \cite{SM}). 
Nano-ARPES data were collected at a photon energy of 60~eV with linear horizontal polarization using a DA30 electron analyser (Scienta-Omicron). 
Samples were cleaved {\it in situ} under ultra-high-vacuum conditions ($\sim$1$\times$$10^{-10}$~mbar) and kept at a temperature of 30~K.
The total instrumental energy resolution including the light, analyser, and thermal broadening were set to be better than 50~meV for spatial mapping, and 35~meV for measuring band dispersions and Fermi surfaces. 
The angular resolution was better than 0.13$^\circ$, estimated by fitting angular distribution curves at the Fermi level ($E_{\rm F}$) using a Voigt function.

Figure~\ref{Fig1}(a) illustrates setup for nano-ARPES experiments, where a submicrometric beam can be provided by utilizing focusing optics in close vicinity to the sample.
For polar angular scans ($\theta$), a deflector scan of the electron analyser lens ($\theta_{a} = \pm 10^\circ$) was used against the detector slit in the vertical direction. 
This does not require changing the geometry between the light and sample, thus ensuring stable illumination of and data acquisition from a microscopic spot of the material. 

Figure~\ref{Fig1}(b) shows a scanning photoemission microscopy (SPEM) image of Y124 obtained by integrating the valence bands (VB) intensity in a fixed energy window [shaded areas in Fig.~\ref{Fig2}(e)] 
and scanning the sample along two in-plane directions ($X$ and $Y$). 
The overall topography shown in Fig.~\ref{Fig1}(b) is consistent with the {\it ex-situ} optical microscope (OM) images [Fig.~\ref{Fig1}(c)]. 
As the cleaved surface of Y124 is generally expected to contain multiple surface terminations due to either CuO-chains or BaO-layer [Figs.~\ref{Fig1}(d)-(f)]. 
In the following, to reveal intrinsic intensity modulations related with different surface terminations, 
we focus on a small area on the sample surface [red square in Fig.~\ref{Fig1}(b)] where height fluctuations are small ($<$1-3~$\mu$m), well below the focal depth of the nano-ARPES setup.

\begin{figure*}
\includegraphics[width=160mm,keepaspectratio]{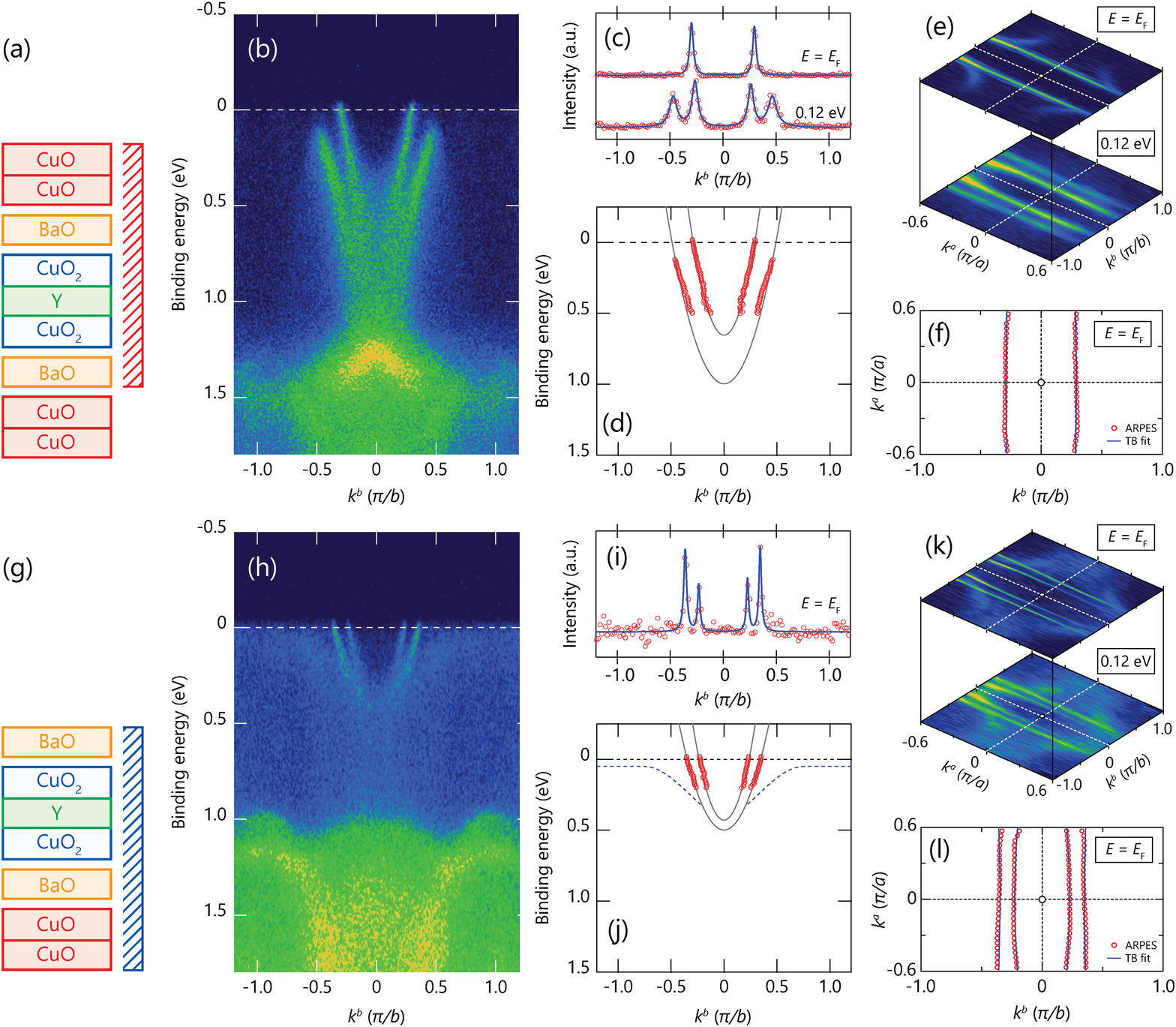}
\caption{
Termination-dependent electronic structures of Y124.
(a) Schematic illustration of region probed by nano-ARPES along the layer stacking direction for the CuO chain termination.
(b) ARPES image on the CuO chain termination taken along the high symmetry Y-$\Gamma$-Y line.
(c) Momentum distribution curves (MDCs) at the Fermi level ($E_{\rm F}$) and a binding energy ($E_{\rm B}$) of 0.12 eV.
(d) MDC derived band dispersions (red circles), where grey lines are calculated band dispersions \cite{Yu91} but shifted in energy to reproduce the experimental Fermi momentum ($k_{\rm F}$).
(e) Constant energy maps at $E_{\rm F}$ as well as at $E_{\rm B}$ = 0.12 eV. 
(f) Experimental $k_{\rm F}$ plots of the CuO chains determined by fitting MDCs, together with a tight-binding fit (see the Supplemental Note~5 \cite{SM}). 
(g)-(l) Same as a to the panels (a)-(f) except for that they are for the BaO surface termination. 
Note that blue dashed line in the panel (j) is eye guides for band dispersions from the CuO$_2$ plane.}
\label{Fig3}
\end{figure*}

In Fig.~\ref{Fig2}, we use nano-ARPES to identify the different surface terminations on the cleaved (001) plane of Y124, and characterize the electronic structure in each case. 
Figure~\ref{Fig2}(a) shows the spatial mapping image of total ARPES intensity near $E_{\rm F}$ along the high-symmetry Y-$\Gamma$-Y line [see Fig.~\ref{Fig2}(c)]. 
As seen in the calculated band dispersions [Fig.~\ref{Fig2}(d)], the electronic structure along the $\Gamma$Y direction near $E_{\rm F}$ consists of two bands, in each case of double CuO chains and bilayer CuO$_2$ planes bands \cite{Yu91}. 
To reveal the separate contributions of the CuO chains and CuO$_2$ planes, we performed an analysis of the spatial mapping data by utilizing narrow integration windows in both energy and momentum. 
First, the integration windows were determined so as to be sensitive to either bands from CuO chains or CuO$_2$ planes as indicated by red and blue boxes in Fig.~\ref{Fig2}(e), 
which yields the two-dimensional intensity maps ($I_{\rm CH}$ and $I_{\rm PL}$, respectively). 
Secondly, we deduced the intensity ratio of the mean deviation between the CuO chains and CuO$_2$ planes as $(I_{\rm CH}-\overline{I_{\rm CH}})/(I_{\rm PL}-\overline{I_{\rm PL}})$,
where the $\overline{I_{\rm CH}}$ and $\overline{I_{\rm PL}}$ are the mean value of $I_{\rm CH}$ and $I_{\rm PL}$, respectively. 
The resulting image shown in Fig.~\ref{Fig2}(d) highlights the dominant signal from to the CuO chain bands in red and the CuO$_2$ plane bands in blue. 
Thirdly, core level spectra in each domain displayed in Fig.~\ref{Fig2}(e) show significant differences, particularly in the binding energies of Ba 5$p$ core levels. 
This chemical shift is unambiguously found in the two types of domains and attributed to different surface terminations of CuO-chains or BaO-layer \cite{Iwasawa18} (see also Supplementary Note~3 \cite{SM}). 
In what follows, we focus on ARPES features that are attributed to the CuO chains but are probed on the two different surface terminations CuO and BaO.

Having thus characterized the CuO-chain or BaO-layer terminated surface regions, 
we perform nano-ARPES measurements on each termination [red region for the CuO-terminated surface and the blue region for the BaO-terminated surface in Fig.~\ref{Fig2}(d)]. 
Owing to its strong surface sensitivity, we can expect the ARPES spectra to be dominated by emission from the top-most surface layer: 
a CuO chains layer for the CuO-terminated surface and a CuO$_2$ plane layer for the BaO-terminated surface as seen in the schematic drawings of the possible stacking layer along the $c$-axis [Figs.~\ref{Fig3}(a), (g)]. 
The BaO layer itself is not expected to contribute to the electronic structure near $E_{\rm F}$ due to its insulating nature. 
Figures~\ref{Fig3}(b), (h) show the ARPES data along the high-symmetry direction Y-$\Gamma$-Y on CuO-chain and BaO-layer terminations, respectively. 
Consistent with previous micro-ARPES results on the CuO-terminated surface \cite{Kondo07,Kondo10}, two strong dispersive features derived from CuO-chains near the surface (${\rm CH_{CuO}^{TS}}$) are observed centred at the $\Gamma$ point. 
The inner-chain band dispersion crosses $E_{\rm F}$, while the outer-chain band does not cross $E_{\rm F}$, but rather reaches only up to a maximum point located at $\sim$0.12 eV below $E_{\rm F}$. 
Similarly, for the BaO-terminated surface, 
two bands exist centred at the $\Gamma$ point besides the broad CuO$_2$ plane bands dispersing from the $\Gamma$ point towards the Y point at $k^a = \pm\pi/a$ [blue dashed line in Fig.~\ref{Fig3}(j)]. 
Here, observed for the first time in our data, we find sharp metallic dispersions for the both states of the double CuO chains. 
It is worth stressing that these observed CuO chains states on the BaO-terminated surface must arise from sub-surface CuO-chains layer (${\rm CH_{BaO}^{SS}}$) beneath the CuO$_2$ planes 
and these states are significantly different from those observed in ${\rm CH_{CuO}^{TS}}$.

The observed distinct behaviours between the ${\rm CH_{CuO}^{TS}}$ and ${\rm CH_{BaO}^{SS}}$ are also visualised in the momentum distribution curves (MDCs) in Figs.~\ref{Fig3}(c), (i), and MDC-derived band dispersions in Figs.~\ref{Fig3}(d), (j). 
Astonishingly, the MDC dispersions (red circles) can be fitted very well by a simple shift of the band dispersions calculated by density functional theory (DFT) \cite{Yu91} (grey lines) along the energy axis as shown in Figs.~\ref{Fig3}(d), (j). 
This implies a small effect of mass renormalization due to electron correlations on either surface termination. 
An analysis of effective masses relative to the free electron mass $m_e$ gives $m^*=0.92~m_e$ for the outer and $m^*=0.49~m_e$ for the inner CuO chain bands, irrespective of surface terminations. 

The constant energy planes shown in Figs.~\ref{Fig3}(e), (k) highlight again the difference between the ${\rm CH_{CuO}^{TS}}$ and ${\rm CH_{BaO}^{SS}}$, 
which can be confirmed not only along the high-symmetry line but also in a wider region across the Brillouin zone. 
Either ${\rm CH_{CuO}^{TS}}$ and ${\rm CH_{BaO}^{SS}}$ are dominant in the maps, while the bands of the CuO$_2$ plane are less visible [Figs.~\ref{Fig3} (e), (k)]. 
We then determine the quasi-1D CuO-chain Fermi surfaces precisely by fitting MDCs as shown in Figs.~\ref{Fig3}(f), (l), where a very small but finite warping along the $k^a$ is observed. 
This is assigned to finite interchain coupling. 
To evaluate the in-plane charge transfer due to interchain and intrachain couplings, we fit the Fermi momentum plots using a simple tight-biding model for one-dimensional chain \cite{Nicholson17} given by 
\begin{equation}
E_k = - 2 t_a \cos (k^a \cdot a) - 2 t_b \cos (k^b \cdot b) - \mu,
\label{Eq1}
\end{equation}
where $\mu$ is the chemical potential, and $t_a$ and $t_b$ are hopping parameters across and along the chain, respectively (see also Supplementary Note~4 \cite{SM}). 
The quasi-1D dispersions can be well reproduced by the model with the parameters summarized in Table~\ref{Table1}. 
We found that the interchain hopping $t_a$ is consistently small compared to the intrachain hopping for all CuO-chain bands, resulting in the large hopping anisotropy given by $|t_{b}/t_{a}|$.

Meanwhile, the tight-binding fit of the Fermi surface also provides an estimation of the carrier concentration. 
We confirmed that the carrier concentration is almost maintained for the CuO-terminated surface while extra hole-doping is present in the case of the BaO-terminated surface (see Supplementary Note~5 \cite{SM}). 
Despite the hole-doping difference, we interpret that the metallic CuO-chain states ${\rm CH_{BaO}^{SS}}$ on the BaO-terminated surface is more representative of bulk properties because 
(1) they are free from exotic surface insulation effects seen in the top-most CuO-chain exposed on the CuO-terminated surface and 
(2) the inner CuO-chain bands on both surface-terminations show a rigid-band shift as well as the similar Fermi surface warping.

Our experimental observations of the metallic ${\rm CH_{BaO}^{SS}}$ states provide important insights into the in-plane charge dynamics on the 1D CuO chains in the Y124 system. 
According to ch-DMFT calculations for the coupled 1D chains \cite{Berthod95}, 
the 1D chain Fermi surface shows a strong anisotropy due to the interchain coupling with an intermediate coupling strength ($t_a^{c1} < t_a < t_a^{c2}$, where $t_a^{c1}$ and $t_a^{c2}$ are some critical hopping parameters). 
This may lead to a formation of Fermi pockets which are disconnected at ($k^a$, $k^b$) = (0, $\pi/2b$) because of the divergence of the self-energy. 
Indeed, the calculated intensity distributions were similar to what were observed in the previous micro-ARPES results \cite{Kondo10}. 
However, this scenario must be inapplicable to the present system because the interchain hopping is found to be critically small ($t_a < 0.03 t_b \ll t_a^{c1}$).
This interpretation is also supported by the experimental observation of the continuous Fermi surface with a finite spectral weights at $E_{\rm F}$ around ($k^a$, $k^b$) = (0, $\pi/2b$) [Figs.~\ref{Fig3}(e), (k)].

The observed strong hopping anisotropy in the CuO chain, on the other hand, should enhance the 1D directionality of the CuO-chain electronic states, 
which may lead to characteristic 1D electronic excitations of a Tomonaga-Luttinger liquid (TLL). 
However, we did not observe the power-law depletion of spectral weight in the vicinity of $E_{\rm F}$ \cite{Voit95}, given by $\rho (\omega) \propto |\omega|^\alpha$. 
Furthermore, we did not observe any steep broadening of quasiparticle lifetime near $E_{\rm F}$ with a linear $\omega$-dependence \cite{Varma96}
expected for a marginal Fermi liquid (FL) nor the significant effective mass enhancement due to electron correlations.
We therefore see that the electrons on CuO-chains behave like 1D Fermi gas along the CuO chain in Y124. 
Meanwhile, we also predict that a longitudinal hopping ($t_c$) may exist. 
An orbital hybridization between the CuO chain $3d_{y^2-z^2}$ orbitals and CuO$_2$ plane $3d_{3z^2-r^2}$ orbitals 
via apical oxygen 2$p_z$ orbitals was predicted based on more bulk sensitive experiments on the analogous Y123 system using x-ray absorption spectroscopy (XAS) and resonant inelastic x-ray scattering (RIXS) \cite{Magnuson14}. 
This interpretation is also in line with the 3D metallic transport properties of Y124 \cite{Hussey97}. 
Further studies on the $t_c$ hopping would be of interest to address this point.

\begin{table}[t]
\caption{Interchain and itrachain hopping parameters ($t_a$ and $t_b$), and hopping anisotropy $|t_b/t_a|$ of the inner and outer CuO$_2$-chain bands (CH$_{\rm inner}$ and CH$_{\rm outer}$) for CuO-terminated and BaO-terminated surfaces,
obtained by fitting tight-binding model to experimental Fermi surfaces. 
}
\label{Table1}
\begin{ruledtabular}
\begin{tabular}{lllll}
&
& $t_a$ [eV]
& $t_b$ [eV]
& $|t_b/t_a|$\\
\hline
CuO
& CH$_{\rm inner}$ 
& 0.003
& 0.20
& 66.7
\\
\hline
\multirow{2}{*}{BaO}
& CH$_{\rm inner}$ 
& 0.004
& 0.12
& 30.0
\\

& CH$_{\rm outer}$ 
& -0.004
& 0.17
& 42.5
\\

\end{tabular}
\end{ruledtabular}
\end{table}

In summary, we present the termination-dependent electronic structure of Y124 using nano-ARPES. 
The observed electronic properties of the CuO chains are quite different between the CuO- and BaO-terminated surfaces. 
The exposed CuO chains (${\rm CH_{CuO}^{TS}}$) on the CuO termination show gapped features reflecting significant surface effects, consistent with the earlier reports using micro-ARPES results \cite{Kondo07,Kondo10}. 
In contrast, we observe the double metallic dispersions for the CuO chains (${\rm CH_{BaO}^{SS}}$) buried underneath the CuO$_2$ plane on the BaO-terminated surface for the first time. 
The uncovered metallic ${\rm CH_{BaO}^{SS}}$ bands on the BaO-terminated surface exhibit a highly 1D Fermi surface 
with a very small warping due to interchain coupling ($ t_a < 0.03 t_b$) as well as a hopping anisotropy between inter- and intra-chain couplings ($|t_b/t_a|>30$). 
Nevertheless, we do not find any signature of neither the 1D TLL nor marginal FL, and the electrons in the CuO-chains do not undergo significant electron correlations as represented by their small effective masses and un-renormalized band width. These findings should be implemented in a future theoretical framework, which may help to understand the high-$T_c$ superconductivity in YBCO cuprate systems.

\begin{acknowledgments}
H.I. thanks Matthew Watson for useful discussions on developing performance of nano-ARPES instruments. 
We thank Diamond Light Source for access to beamline I05 (proposals no. NT21083) that contributed to the results presented here.
\end{acknowledgments}

\end{document}


\title{Supplemental Material:\\
Buried double CuO chains in YBa$_2$Cu$_4$O$_8$ uncovered by nano-ARPES}

\author{Hideaki Iwasawa}
\thanks{Correspondence and requests for materials should be addressed to H.I. (h-iwasawa@hiroshima-u.ac.jp)}

\affiliation{Diamond Light Source, Harwell Science and Innovation Campus, Didcot OX11 0DE, UK}
\affiliation{Graduate School of Science, Hiroshima University, Hiroshima 739-8526, Japan}

\author{Pavel Dudin}
\affiliation{Diamond Light Source, Harwell Science and Innovation Campus, Didcot OX11 0DE, UK}

\author{Kyosuke Inui}
\author{Takahiko Masui}
\affiliation{Department of Physics, Kindai University, Higashi-Osaka 577-8502, Japan}

\author{Timur K. Kim}
\author{Cephise Cacho}
\affiliation{Diamond Light Source, Harwell Science and Innovation Campus, Didcot OX11 0DE, UK}

\author{Moritz Hoesch}
\affiliation{Diamond Light Source, Harwell Science and Innovation Campus, Didcot OX11 0DE, UK}
\affiliation{Deutsches Elektronen-Synchrotron (DESY), Photon Science, Hamburg 22607, Germany}

\maketitle

\section{\large\bf Supplementary Note 1: Nano-ARPES system}
\par\vspace{5pt}\noindent
The experiments presented here were performed using the newly developed nano-ARPES instruments at beamline I05 of the Diamond Light Source.
Except for details on the nano-ARPES end-station as mentioned below, a description of the beamline as well as the branched high-resolution instruments are found elsewhere \cite{Hoesch17}.
The beamline and end-station are designed to utilize rich functionalities such as variable photon energies (60-100 eV), polarisations (linear horizontal/vertical and left/right hands circular), and sample temperatures (down to 25 K).
To perform ARPES with a small spot effectively, a deflector scan ($\pm$10~deg) of the electron analyser (DA30, Scienta-Omicron)
or the rotation of the analyser itself (-5 to +30~deg) are available while keeping a spatial relationship between light and sample. 
In addition, the polar angles (-10 to 45~deg) and azimuthal angles ($\pm$15~deg) of the sample can be varied by a sample goniometer.
The focusing of the light was done by using the first diffraction order of Fresnel zone plate (FZP).
The zeroth diffraction order and higher diffractions orders were removed by an order sorting aperture (OSA) sitting between the sample stage and FZP.
The practical spatial resolution of the beamline was evaluated to be better than 500~nm for all the available photon energies via knife edge profiles as described in the next section.

\par\vspace{10pt}
\section{\large\bf Supplementary Note 2: Evaluation of spatial resolution}
\par
\begin{figure*}[t]
\begin{center}
\includegraphics[width=160mm,keepaspectratio]{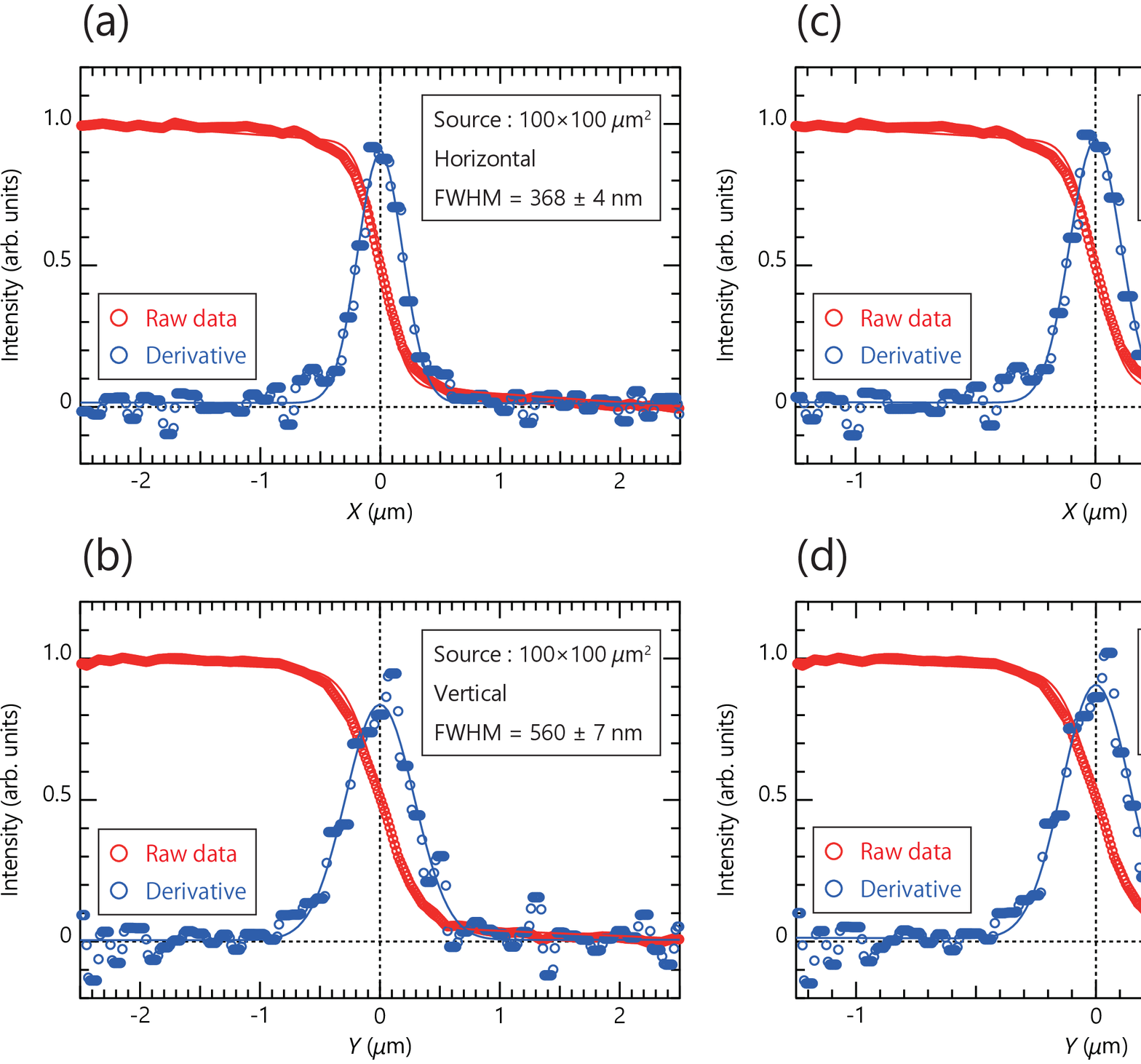}
\caption{
Evaluation of spatial resolution using a gold knife edge profile collected with 60~eV, linear polarization.
(a), (b) Horizontal and vertical knife edge profiles with the $100 \times 100~\mu$m$^2$ source, respectively. 
(c), (d) Same as (a)-(b) except for the source is $20 \times 20~\mu$m$^2$. 
} 
\label{FigS1}
\end{center}
\end{figure*}

\vspace{5pt}\noindent
Here we present the evaluation of the spatial resolution of present nano-ARPES instruments at I05 beamline, Diamond Light Source, using a gold knife edge profile taken with 60~eV (linear horizontal polarization). 
The spatial resolution was evaluated by a simultaneous curve fitting on a raw knife edge profile [$I(X)$ and $I(Y)$ along the horizontal and vertical directions, respectively] 
and its derivative [$dI(X) / dX$ and $dI(Y) / dY$] using a Step function and Gaussian function, respectively. 
Those actual forms used were: 
\begin{equation}
I_{\rm Step} (t) = 
\begin{cases}
(a_1 + b_1 t + c_1) * g_1 (t, d, w) & (t \leq d) \\
(a_1 + b_1 t) * g_1 (t, d, w) & (t > d)
\end{cases}
\label{EqS1}
\end{equation}
and
\begin{equation}
I_{\rm Gauss} (t) = a_2 + b_2 t + g_2 (t, d, w) 
\label{EqS2}
\end{equation}
where $t=X$ or $Y$, $a_{1,2}$ and $b_{1,2}$ are coefficients of a linear background, $c_1$ is a step height when the $t$ is smaller than an edge position $d$, 
and $g_1 (t,d,w)$ is a Gaussian function convoluting the spatial resolution $w$, and $g_2 (t,d,w)$ is a Gaussian function centered at the $d$.
As seen in the equations, the $w$ and $d$ are the linked fitting parameters.
Figures~S1(a), (b) show the raw edge profiles (red circles) and those derivative (blue circles) in case of the source size of $100 \times 100$~$\mu$m$^2$, where the fitting curves (red and blue lines) are also plotted.
 The horizontal and vertical broadenings in full width half maximum (FWHM) were determined as $368 \pm 4$~nm and $560 \pm 7$~nm, respectively, 
and we thus obtained $454 \pm 12$~nm$^2$ for the spatial resolution from the square root of those product. 
The superior spatial resolution can be achieved by reducing the source size down to $20 \times 20$~$\mu$m$^2$ as shown in Figs.~\ref{FigS1}(c), (d). 
In the same manner as above, we obtained the spatial resolution of $240 \pm 12$~nm$^2$ with the horizontal and vertical broadenings of $209 \pm 2$~nm and $276 \pm 4$~nm, respectively. 

\par\vspace{10pt}
\section{\large\bf Supplementary Note 3: Surface termination dependent shallow core level spectra}
\par\vspace{5pt}\noindent
We show the angle-integrated energy distribution curves (EDCs) measured on the CuO-terminated (red curve) and BaO-terminated (blue curve) surfaces in Fig.~2(e).
Those spectral peaks can be assigned to the shallow core levels of constituent elements of YaBa$_2$Cu$_4$O$_8$ (Y124) similar to those reported in an analogue YBa$_2$Cu$_3$O$_7$ system (Y123) \cite{Thiess07,Thiess15}, 
though the Ba 5$s$ and Y 4$p$ are not visible on the CuO-terminated surface. 
Notably, the Ba 5$p_{1/2}$ and 5$p_{3/2}$ shallow core levels show a clear energy shift ($\sim$1.4~eV) between the two surface terminations. 
The observed shift is larger than the surface core level shifts, which is typically $\sim$1~eV or less. 
The reported surface core level shift on the Ba 5$p$ peaks is $\sim$0.9~eV in the analogous Y123 \cite{Maiti09}.
It is thus quite difficult to expect that only termination differences on the topmost surface with/without the coverage of the CuO chains give rises to such large energy shifts. 
Alternatively, it is more reasonable to consider that the observed shift has a chemical origin. Indeed, negative shifts in the binding energy of the Ba 5$p$ core levels were observed by oxygen adsorption on a Ba surface \cite{Vlachos06}.
Such chemical shifts are qualitatively consistent with what we observed here, that is, the shallower (deeper) peak is demonstrating that each measured position is dominantly terminated by a BaO (CuO) layer.
This assignment is qualitatively supported by the observations that the peak intensity of Ba 5$s$ (5$p$) and Y 4$p$ is greatly (slightly) suppressed for the CuO terminated surface compared with the BaO terminated surface,
in consistent with the natural expectation from the $c$-axis layer stacking (BaO and Y layers become deeper for the CuO terminated surface).

\par\vspace{10pt}
\section{\large\bf Supplementary Note 4: Tight-binding fit}
\par\vspace{5pt}\noindent
We employed a simple tight binding model for one-dimensional chain given by Eq.~1 \cite{Nicholson17}. 
Considering the $\Gamma$ point ($k^a=k^b=0$ and $E_k=E_\Gamma$) and the Fermi points ($k^b=k_{\rm F}^b$)
at the Fermi energy with $k^a=0$, we have respectively
\begin{equation}
E_\Gamma = - 2 t_a - 2 t_b - \mu,
\label{Eq2}
\end{equation}
\begin{equation}
0 = - 2 t_a - 2 t_b \cos (k_{\rm F}^b \cdot b) - \mu.
\label{Eq3}
\end{equation}
Then, these two equations lead to
\begin{equation}
t_b = - \frac{E_\Gamma}{2 [1+\cos(k_{\rm F}^b \cdot b)]},
\label{Eq4}
\end{equation}
\begin{equation}
\mu = - 2 t_a - 2 t_b \cos (k_{\rm F}^b \cdot b).
\label{Eq5}
\end{equation}
Consequently, a fitting function for the Fermi momentum plot can be obtained as
\begin{equation}
k^b = \frac{1}{b} \cos ^{-1} 
\left[ 
\frac{t_a}{t_b}
\left\{
1 - \cos (k^a \cdot \pi)
\right\}
+ \cos (k_{\rm F}^b \cdot \pi)
\right].
\label{Eq6}
\end{equation}
Since the $E_\Gamma$ and $k_{\rm F}^b$ were experimentally determined, $t_b$ becomes a constant value, and hence, the $t_a$ is the only the fitting parameter.

\begin{figure*}[b]
\begin{center}
\includegraphics[width=160mm,keepaspectratio]{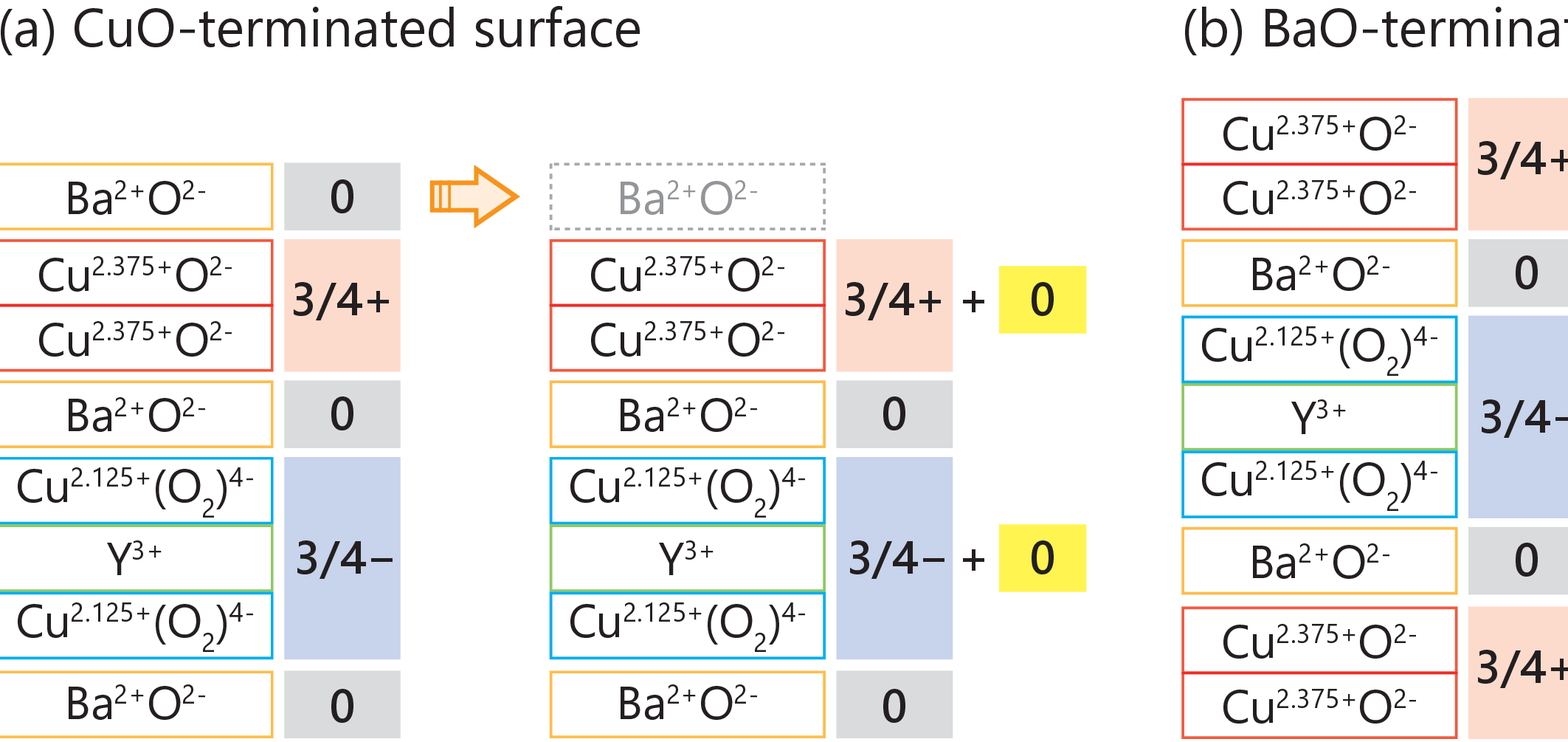}
\caption{
Schematic illustrations of electronic reconstructions on the cleaved surface of YBa$_2$Cu$_4$O$_8$.
Model electronic configurations before (left) and after (right) cleaving for (a) CuO-terminated surfaces and (b) BaO-terminated surfaces.
} 
\label{FigS2}
\end{center}
\end{figure*}

\par\vspace{10pt}
\section{\large\bf Supplementary Note 5: Surface termination dependent electronic reconstruction}
\par\vspace{5pt}\noindent
From the fit of the tight-binding model to the Fermi surface plot [Figs.~3(f), (l)], 
we can evaluate the doping level of the top-most CuO-chains on the CuO-terminated surface (${\rm CH_{CuO}^{TS}}$) and sub-surface CuO-chains on the BaO terminated surfaces (${\rm CH_{BaO}^{SS}}$) \cite{Comment1}. 
The Fermi surface area, relative to the Brillouin zone area ($4\pi^2/ab$), 
was determined to be $S_{\rm ave} = 31.5\%$ and 22.9$\%$ for the ${\rm CH_{CuO}^{TS}}$ and ${\rm CH_{BaO}^{SS}}$ states, respectively, from the average of the inner and outer bands. 
Thus, the corresponding numbers of electrons $n$ are counted as 0.63 and 0.46 for CuO- and BaO-terminated surfaces, indicating the hole-rich states (overdoped) on the BaO-terminated surface. 
This charge re-distribution can be explained as below by a surface-termination dependent electronic reconstruction model as we proposed in an analogous Y123 system \cite{Iwasawa18}.

The surface termination dependent electronic reconstruction across the cleaving is illustrated on each surface termination in Fig.~\ref{FigS2}.
Assuming the nominal hope doping $p$ = 0.125 into the CuO$_2$ planes \cite{Yelland08}, the nominal charges become +2.375$e$ per Cu-chain with $n$ = 0.625 to hold an overall charge neutrality [left-side of Figs.~\ref{FigS2}(a), \ref{FigS2}(b)]. 
In the case of the CuO-terminated surface, the charge environment around the top-most CuO-chains is not varied before and after cleaving [Fig.~\ref{FigS2}(a)], resulting in no charge transfer.
However, in the case of the BaO-terminated surface, the removal of the double Cu-chains across cleaving results in a charge imbalance with respect to the topmost BaO layer [Fig.~\ref{FigS2}(b)].
As a result, the removed charge of 3/4+ has to be compensated from neighboring atoms to maintain the overall charge neutrality within a top-most unit cell.
Considering a uniform charge redistribution (3/8+ each) within a unit cell, 
the average charges become +2.3125$e$ in the CuO$_2$-plane site with the hole doping $p$ = 0.31, and +2.5625$e$ in the subsurface Cu-chain site [right-side of Fig.~\ref{FigS2}(b)] with $n$ = 0.4375.

Consequently, we obtained the qualitative agreement between the ARPES and model estimation of the number of electrons in the Cu-chain site 
(ARPES: $n$ = 0.63 and 0.46, Model: $n$ = 0.625 and 0.4375, for CuO- and BaO-terminated surfaces, respectively).
We note that the overdoped CuO$_2$-plane states on the BaO-terminated surface are qualitatively consistent between the present model ($p$=0.31) and previous ARPES results ($p$=0.23) \cite{Kondo07}.
The quantitative difference might be introduced by a further charge distribution occurred from the model electronic configurations [right panels of Figs.~\ref{FigS2}(a), \ref{FigS2}(b)] 
as the surface insulation was induced into the top-most CuO-chain on the CuO-terminated surface. 
Further studies including detailed chemical analysis may allow to fully uncover the surface electronic reconstruction in Y124.

\par\vspace{10pt}
\section{\large\bf Supplementary References}
\par